# Bypassing the structural bottleneck in the ultrafast melting of electronic order

L. X. Yang[1,2,3,*], G. Rohde[1], K. Hanff[1], A. Stange[1], R. Xiong[4], J. Shi[4], M. Bauer[1], K. Rossnagel[1,5,†]

[1]*State Key Laboratory of Low Dimensional Quantum Physics, Department of Physics, Tsinghua University, Beijing 100084, P. R. China*

[2]*Institut für Experimentelle und Angewandte Physik, Christian-Albrechts-Universität zu Kiel, 24098 Kiel, Germany*

[3]*Frontier Science Center for Quantum Information, Beijing 100084, P. R. China*

[4]*Department of Physics, Wuhan University, Wuhan 430072, P. R. China*

[5]*Ruprecht-Haensel-Labor, Deutsches Elektronen-Synchrotron DESY, 22607 Hamburg, Germany*

*lxyang@mail.tsinghua.edu.cn

†rossnagel@physik.uni-kiel.de

**The emergent properties of quantum materials, such as symmetry-broken phases and associated spectral gaps, can be effectively manipulated by ultrashort photon pulses. Impulsive optical excitation generally results in a complex non-equilibrium electron and lattice dynamics that involves multiple processes on distinct timescales, and a common conception is that for times shorter than about 100 fs the gap in the electronic spectrum is not seriously affected by lattice vibrations. Here, we directly monitor the photo-induced collapse of the spectral gap in a canonical charge-density-wave material, the blue bronze $Rb_{0.3}MoO_3$. We find that ultra- fast (∼60 fs) vibrational disordering due to efficient hot-electron energy dissipation quenches the gap significantly faster than the typical structural bottleneck time corresponding**

**to one half-cycle oscillation (≈315 fs) of the coherent charge-density-wave amplitude mode. This result not only demonstrates the importance of incoherent lattice motion in the photo-induced quenching of electronic order, but also resolves the perennial debate about the nature of the spectral gap in a coupled electron-lattice system.**

**Introduction**

In quantum materials, the coupling between the electronic and lattice degrees of freedom is often strong so that changes in the electronic properties can lead to structural distortions and vice versa. Charge-density-wave (CDW) bearing materials provide classic examples of such intertwining[1,2], as well as of the resulting controversies as to whether the transition to a charge- and lattice-modulated state and the emerging energy gap (Fig. 1a) are predominantly due to electron-phonon coupling or electron-electron interactions[3–6]. One allure of femtosecond time-resolved pump-probe techniques is that they can provide novel insights into this problem via temporal discrimination of electronically and lattice-driven processes[7–29]. Specifically, after impulsive photo-excitation, the electronic and lattice components of CDW order and of the energy gap can be expected to decouple because of significantly different response times. A common scenario is a multi-step quench of the CDW state in which the electronic order is suppressed in less than 100 fs, while the lattice distortion is coherently relaxed on CDW amplitude-mode vibrational timescales of a few 100 fs, and incoherent lattice disordering takes place on a 1-ps timescale through electron-phonon thermalization[8,9,11–13,16,17,20,23]. *Coherent* lattice motion can thus be understood as a speed-limiting structural bottleneck for the quenching of combined charge and lattice order[8,10,13,16,17,21,27,29], although a fully *incoherent* structural transition pathway on the timescale of a single phonon oscillation is also possible[30].

Here, we present direct spectroscopic evidence that *incoherent* lattice motion can indeed cause complete CDW gap quenching well before *coherent* lattice motion effectively modulates the gap. Our key observation is that the initial gap quenching correlates with a surprisingly fast hot-electron energy relaxation implying a quasi-instantaneous generation of high-frequency lattice fluctuations. These photo-induced non-thermal fluctuations can rapidly fill in the gap and smear the gap edge similar to the effects of thermal lattice fluctuations[31–33] (Fig. 1b).

Our experimental tool is time- and angle-resolved photoemission spectroscopy[9, 10, 15, 16] (trARPES) using near-infrared pump and extreme ultraviolet probe pulses[34] (Fig. 1c). The technique provides a direct momentum-resolved view on energy-gap dynamics with an effective time resolution of a few 10 fs, short enough to reveal and separate the responses of the electronic and lattice components of the order[15–17,21]. The material is the quasi-one-dimensional blue bronze $Rb_{0.3}MoO_3$ (ref. 35), in which double chains of $MoO_6$ octahedra run along the crystallographic *b* direction (Fig. 1c). This prototype CDW system undergoes a transition at $T_{CDW} \approx 180$ K, and the saturation value of the energy gap ($2\Delta_0$) is about 120 meV (ref. 5).

**Results**

**Spectral intensity dynamics**. Figure 1d shows a representative ARPES band map measured without optical excitation below $T_{CDW}$ along the one-dimensional chain ($\Gamma$-$Y$) direction. Consistent with high-resolution ARPES data[5,6] and the results of band structure calculations[36,37], the two relevant bands near the Fermi level ($E_F$) are resolved: the weakly intense and weakly dispersive anti-bonding (AB) band and the strongly intense and strongly dispersive bonding (B) band, which we will focus on in the following.

The photo-induced temporal evolution of the spectrum around the Fermi wavevector of the B band is shown in Fig. 1e and f for equilibrium sample temperatures below and above $T_{CDW}$, respectively. Upon excitation of the CDW state ($T < T_{CDW}$), the spectral weight peak associated with the B band is first suppressed in intensity and then shifts toward $E_F$. For the normal state ($T > T_{CDW}$), by contrast, no such peak shift is observed under otherwise identical measurement conditions. This distinct temperature-dependent spectrotemporal response is further corroborated by the comparison of energy distribution curves (EDCs) at selected pump-probe time delays shown in Fig. 1g (see also Supplementary Fig. S1).

The time- and energy-dependent ARPES intensity difference map in Fig. 2a reveals how the spectral weight dynamics in the CDW state proceeds in characteristic energy windows above and around $E_F$. Figure 2b compares the corresponding transient ARPES intensity integrated over portions of these distinct intervals (see also Supplementary Fig. S2).

Far above $E_F$ and the gap (≈0.3–0.6 eV), the quasi-instantaneous rise (∼30 fs, limited by the experimental time resolution) and rapid exponential decay (time constant of 30 ± 5 fs) of the spectral weight reflect the creation and relaxation of hot electrons at high energies, respectively. Directly above the gap edge (≈0.06–0.3 eV), the initial rise is still quasi-instantaneous, but the electron relaxation considerably slows down and shows a bi-exponential decay (time constants of 120 ± 20 fs and 1900 ± 200 fs). Figure 2c shows that this electron energy-dependent intensity relaxation is reflected in bi-exponential decay (time constants $\tau_{e,1} = 45 \pm 6$ fs, $\tau_{e,2} = 580 \pm 150$ fs) of the transient total electron energy as approximated by the first moment of the spectral weight distribution above $E_F$, $E_t = \int_0^{0.85\ eV} E I_B(E,t) dE$. The electron density, on the other hand, which is roughly proportional to the transient integrated spectral weight, $N_t = \int_0^{0.85\ eV} I_B(E,t) dE$, displays single exponential decay (time constant of 330 ± 50 fs). The observed dynamics is consistent with rapid, substantial hot-electron energy relaxation through phonon emission followed by slower electron-phonon thermalization and recombination of low-energy electrons across the gap region[38].

Remarkably, qualitatively different dynamics is detected in the CDW gap region (≈−0.06–1.6 eV). The initial increase in intensity is prolonged to about 100 fs (sigmoidal time constant $\tau_\Delta = 60 \pm 10$ fs, dotted vertical line in Fig. 2b) and followed by a characteristic intensity plateau[17,21], which extends until about 280 fs after excitation (dash-dotted vertical line in Fig. 2b), before relaxation sets in (with an exponential time constant of 120 ± 40 fs). Such transient behavior is reminiscent of the typical two-step melting of CDW order in which a rapid electronic quench is superimposed on a slower damped coherent lattice relaxation[9,16,17,21]. Indeed, the end time of the plateau corresponds to the typical structural bottleneck time, $\tau_A/2$, where $\tau_A$ is the oscillation period of the CDW amplitude mode ($\tau_A/2 \approx 315$ fs for $Rb_{0.3}MoO_3$ (ref. 39), see also Supplementary Fig. S3). The intriguing observation, however, is that the filling-in of the gap is much faster and appears to be correlated, not with the photo-induced heating of the electrons, as expected for a purely electronic process, but with the initial fast electron energy relaxation (see orange, blue, and dotted vertical lines in Fig. 2b). In the Supplementary Information, this novel aspect to the generic two-timescale spectral response is brought out in a direct comparison with a layered CDW material (Supplementary Fig. S4).

**Spectral shape dynamics**. To further quantify and elucidate the gap dynamics, we plot in Fig. 3 the results of a lineshape analysis in which the time-dependent quasiparticle peak at the lower gap edge was approximated with a Lorentzian on a constant background, multiplied by a Fermi-Dirac function and convoluted with a Gaussian representing the experimental energy resolution (Supplementary Fig. S5). At the available experimental energy resolution (250 meV), the model can capture the dynamics of three characteristic spectral parameters—the broadening of the electron energy distribution ($k_B \Delta T_e$) and the shift ($\Delta E_{shift}$) and broadening ($\Delta E_{broad}$) of the quasiparticle peak—irrespective of the fact that the distribution function may be non-thermal and the spectral function non-Lorentzian[5,6,33] (Supplementary Fig. S5). At a crude level, the three parameters may be regarded as measures for the electron temperature, reduction of the energy gap, and quasiparticle scattering rate, respectively.

Figure 3a compares the extracted time dependencies of $k_B \Delta T_e$, $\Delta E_{shift}$ and $\Delta E_{broad}$ for a data set collected at an incident pump fluence of $F = 0.59$ mJcm$^{-2}$. The transients of $k_B \Delta T_e$ and $\Delta E_{shift}$ reproduce the temporal evolutions of $E_t$, the total electron energy above $E_F$ (Fig. 2c), and the spectral weight in the gap region (Fig. 2b), respectively. The quasi-instantaneous (∼30 fs) increase of $k_B \Delta T_e$ to a value corresponding to about 1100 K is followed by a bi-exponential decay ($\tau_{e,1} = 40 \pm 5$ fs, $\tau_{e,2} = 2300 \pm 150$ fs), whereas the transient $\Delta E_{shift}$ displays a slower rise to 65 meV with a sigmoidal time constant $\tau_{shift} = 60 \pm 10$ fs, stationary behavior between 100 fs and 280 fs, and then exponential decay (with a time constant of $220 \pm 60$ fs). We note that the transient peak broadening $\Delta E_{broad}$ does not show plateau-like behavior, but exhibits a sigmoidal rise with a time constant identical to the one of the peak shift ($\tau_{broad} = 60 \pm 18$ fs) as well as similar exponential decay (with a time constant of $480 \pm 80$ fs).

The correlation between the extracted spectral shift $\Delta E_{shift}$ and broadening $\Delta E_{broad}$ is also seen in the temporal evolution of the corresponding transients at a lower pump fluence (Fig. 3b), and in the fluence-dependent maximum shift and broadening (Fig. 3c, see also Supplementary Fig. S6). The maximum $\Delta E_{shift}$ and $\Delta E_{broad}$ values both display saturation behavior at a critical incident fluence of about 0.6 mJcm$^{-2}$. Notably, the saturation value of the peak shift is about 65 meV, consistent with the half-size of the equilibrium CDW gap ($\Delta_0$) well below $T_{CDW}$ (ref. 5), indicating that for excitation densities larger than the critical fluence the CDW gap is completely quenched[18]. The value of the saturation fluence is in rough agreement with the one reported previously for the complete suppression of the electronic component of CDW order[11].

**Discussion**

The measured spectral weight and shape dynamics thus provide direct evidence for a photo-induced melting of the CDW gap that is, first, delayed with respect to hot carrier generation, but significantly faster than coherent lattice motion ($\tau_\Delta \ll \tau_A/2$) and, second, correlated with the initial fast relaxation of the electron energy (above the gap) and an increase of the quasiparticle scattering rate (below the gap), as quantitatively manifested in the match of time constants: $\tau_\Delta \approx \tau_{e,1} \approx \tau_{broad}$. This correlation between characteristic electronic processes around $E_F$, as directly measured by trARPES, is the central result of this work. A natural explanation is that the suppression of the gap and concomitant reduction of the quasiparticle lifetime are driven by electron-phonon scattering, i.e., by the ultrafast, hot electron-induced generation of vibrational disorder. Indeed, lattice fluctuations are well known to have a strong effect on the electronic properties of low-dimensional CDW systems: They tend to give rise to a pseudogap with filled-in gap states and a smeared and shifted gap edge[31,33] (Fig. 1b), and they can cause strong electron scattering[2]. The ultrafast melting of the CDW gap in essence corresponds to the well-known suppression of the CDW peak due to phonon-induced disorder in ultrafast diffraction[13]. The remarkable aspects are the speed and dominance over the coherent CDW amplitude-mode vibration in the present case. Two types of strongly coupled phonon modes, with vibrational periods short enough to be compatible with the observed fast disordering, are likely to be important: high-frequency (15 THz $\leq f < 2\Delta_0/h \approx$ 29 THz) "phase phonons", which are coupled to oscillations in the phase of the electronic order parameter[32,40,41], and the 28 THz (940 cm$^{-1}$) Mo-O stretching mode, which is sensitive to the CDW transition[42,43]. The elucidation of the specific underlying lattice dynamics, however, is beyond the capabilities of trARPES and calls for a direct probing of phonon couplings and populations by ultrafast diffuse scattering techniques[44]. We finally speculate that the rapid atomic disordering due to mode-selective emission of (high-frequenncy) CDW-coupled phonons by hot electrons, as uncovered here, may be a more generic phenomenon[13,45] than the recently demonstrated collective launch of incoherent phonons driven by an ultrafast change to a highly anisotropic, flat lattice potential[30].

Overall, the spectral weight dynamics revealed for $Rb_{0.3}MoO_3$ appears to be driven by strong electron-phonon coupling. This implies that, in the molybdenum blue bronzes, electron-phonon interactions[5] rather than electron-electron interactions[6] provide the dominant contribution to the spectral properties near $E_F$. More generally, in establishing a novel incoherent pathway for the ultrafast melting of electronic order, our trARPES results highlight the dominant role that lattice fluctuations can play in the sub-100-fs dynamics of low-dimensional quantum materials.

**Methods**

**Sample preparation.** Single crystals of $Rb_{0.3}MoO_3$ were grown by the molten salt electrolytic reduction method[35]. High-purity $Rb_2CO_3$ (99.5%, Ruike) and $MoO_3$ (99.0%, Ruike) were used as raw materials in a ratio of 1:4.3. Typical samples had a thin, narrow, approximately rectangular shape and a size of about $1\times3\times8$ mm$^3$. The long edge of the samples was along the quasi-one-dimensional chain direction. The structural characterization was done by selected area electron diffraction and x-ray diffraction. Samples were cleaved *in situ* at 95 K.

**Time- and angle-resolved photoemission spectroscopy.** Pump-probe ARPES measurements were performed using a high-harmonic-generation source and a hemispherical electron analyzer (SPECS Phoibos 150). The effective time and energy resolutions were 35 fs and 250 meV, mainly governed by the duration of the near-infrared (790 nm) pump pulses and the bandwidth of the extreme ultra-violet (22.1 eV) probe pulses, respectively. The momentum resolution was estimated to be 0.1 Å$^{-1}$. The basic light source was a 8-kHz Ti:sapphire amplifier system (KMLabs Dragon pumped by a Lee laser) delivering near-infrared pulses at 1.2 mJ pulse energy and 30 fs (FWHM) pulse duration. The extreme ultraviolet pulses with an estimated pulse duration of less than 10 fs were generated in an argon-filled hollow-fiber waveguide (KMLabs XUUS) pumped with the second harmonic of the fundamental beam at 390 nm. The size of the pump beam was determined to be 365×425 $\mu$m$^2$ by scanning a pin-hole structure at the sample position, and the size of the probe beam was kept sufficiently small to guarantee a uniform excitation.

**Acknowledgements** We thank Holger Fehske and Alexander Kemper for helpful discussions. This work was supported by the German Research Foundation (DFG) through project BA 2177/9-1. L.X.Y. is grateful to the Alexander von Humboldt foundation and the China Postdoctoral Science Foundation for support.

**Author Contributions** L.X.Y. and K.R. conceived the experiments. Time-resolved ARPES measurements were performed by L.X.Y., G.R. and K.H. with support from A.S. Single crystals were grown and characterized by R.X. and J.S. Data analysis and manuscript preparation were done by L.X.Y., M.B. and K.R., with input from all authors.

**Competing Interests** The authors declare no competing financial interests.

**Correspondence** Correspondence and request for materials should be addressed to L.X.Y. (lxyang@tsinghua.edu.cn) and K.R. (rossnagel@physik.uni-kiel.de).

**Figure 1 | Electronic structure and dynamics of the quasi-one-dimensional CDW compound $Rb_{0.3}MoO_3$. a,b**, Schematic illustration of coherent lattice-vibration-induced closing of the CDW gap in the band structure $E(k_\parallel)$ (**a**) and incoherent lattice-fluctuation-induced filling-in of the gap in the density of states $D(E)$ (**b**) (dashed lines, yellow shading). Solid lines and blue shading indicate the characteristic signatures of the canonical CDW state: the opening of the energy gap $2\Delta_0$, the modulation of the conduction electron density, and the associated periodic lattice distortion with lattice periodicity $2a$. **c**, Sketch of the experimental setup and crystal structure of $Rb_{0.3}MoO_3$. **d**, ARPES band map of the unpumped sample along the crystallographic $b$ ($\Gamma$-$Y$) direction ($hv = 22.1$ eV, $T = 95$ K) revealing the dispersion of the bonding (B) and anti-bonding (AB) bands (dashed lines). **e,f**, Time-dependent ARPES spectra taken around the Fermi wavevector of the B band at equilibrium sample temperatures below ($T = 95$ K) (**e**) and above ($T = 220$ K) (**f**) the CDW transition ($F = 0.59$ mJcm$^{-2}$). **g**, Energy distribution curves extracted from the data in **e** and **f** at selected pump-probe time delays. The temporal spectral response of the CDW state is distinct from the one of the normal state.

**Figure 2 | Ultrafast spectral weight dynamics in the CDW state. a**, Energy-versus-time map of the photo-induced change of the ARPES intensity near the Fermi wavevector of the B band ($T = 95$ K, $F = 0.59$ mJcm$^{-2}$). Dashed horizontal lines mark spectral regions with distinct dynamics, whereas the dotted and the dash-dotted vertical lines indicate the two characteristic timescales of the dynamics in the gap region. The filling-in of spectral

weight in the gap ($|E - E_F| \leq 60$ meV) happens on a similar sub-100-fs timescale (dotted vertical line) as the relaxation of the photo-induced spectral weight at high energies ($E - E_F \geq 300$ meV), which is significantly faster than the timescale set by the structural bottleneck (dash-dotted vertical line). **b**, Time- and spectral region-dependent ARPES intensity near the Fermi wavevectors of the B and AB band with fits (solid lines). **c**, Time-dependent zeroth ($N_t$) and first moment ($E_t$) of the B-band ARPES intensity distribution above the Fermi level with fits (solid lines). The quantities $N_t$ and $E_t$ are measures for the number and total energy of the hot electrons above $E_F$, respectively.

**Figure 3 | Ultrafast spectral function dynamics in the CDW state. a,b**, Temporal evolution of the width of the leading edge ($k_B \Delta T_e$) and the broadening ($\Delta E_{broad}$) and shift ($\Delta E_{shift}$) of the spectral function peak, measured at $T = 95$ K and pump fluences of 0.59 mJcm$^{-2}$ (**a**) and 0.23 mJcm$^{-2}$ (**b**). Fits to the data are displayed as solid lines. The dotted and dash-dotted vertical lines indicate characteristic timescales of the transient spectral shift. The timescales for the initial spectral shift and broadening are similar (dotted vertical lines). **c**, Maximum spectral shift and maximum spectral broadening as a function of pump fluence. The data corroborate the correlation between spectral shift and broadening. Error bars represent standard uncertainties of the fitted parameters.

# Figure 1

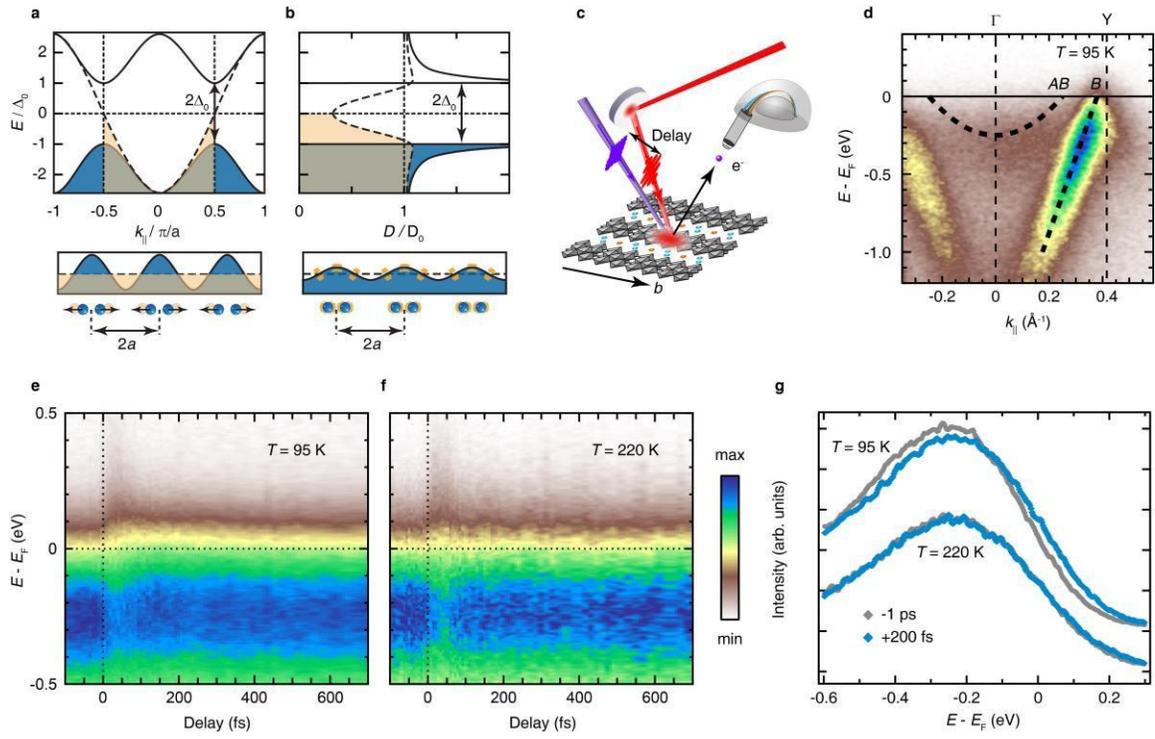

**Figure 2**

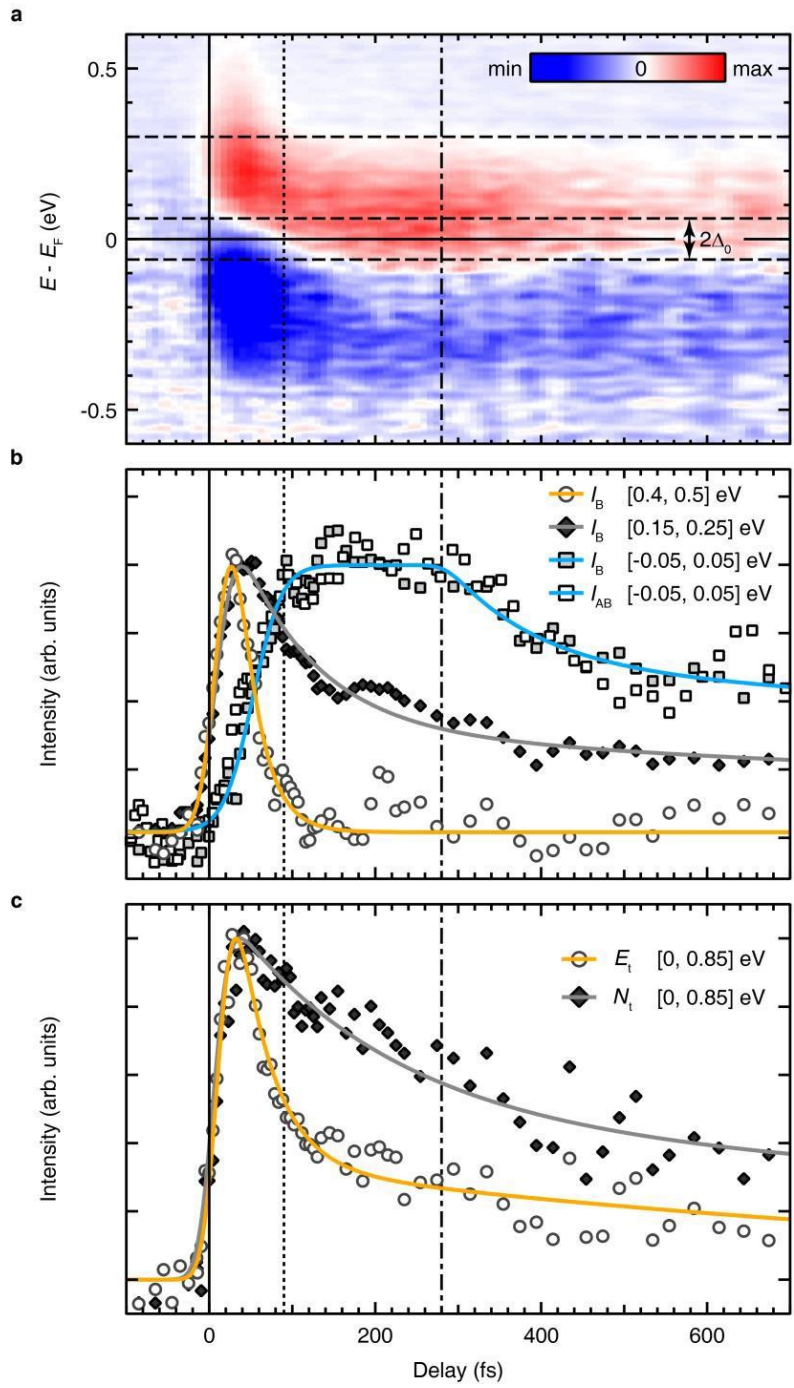

**Figure 3**

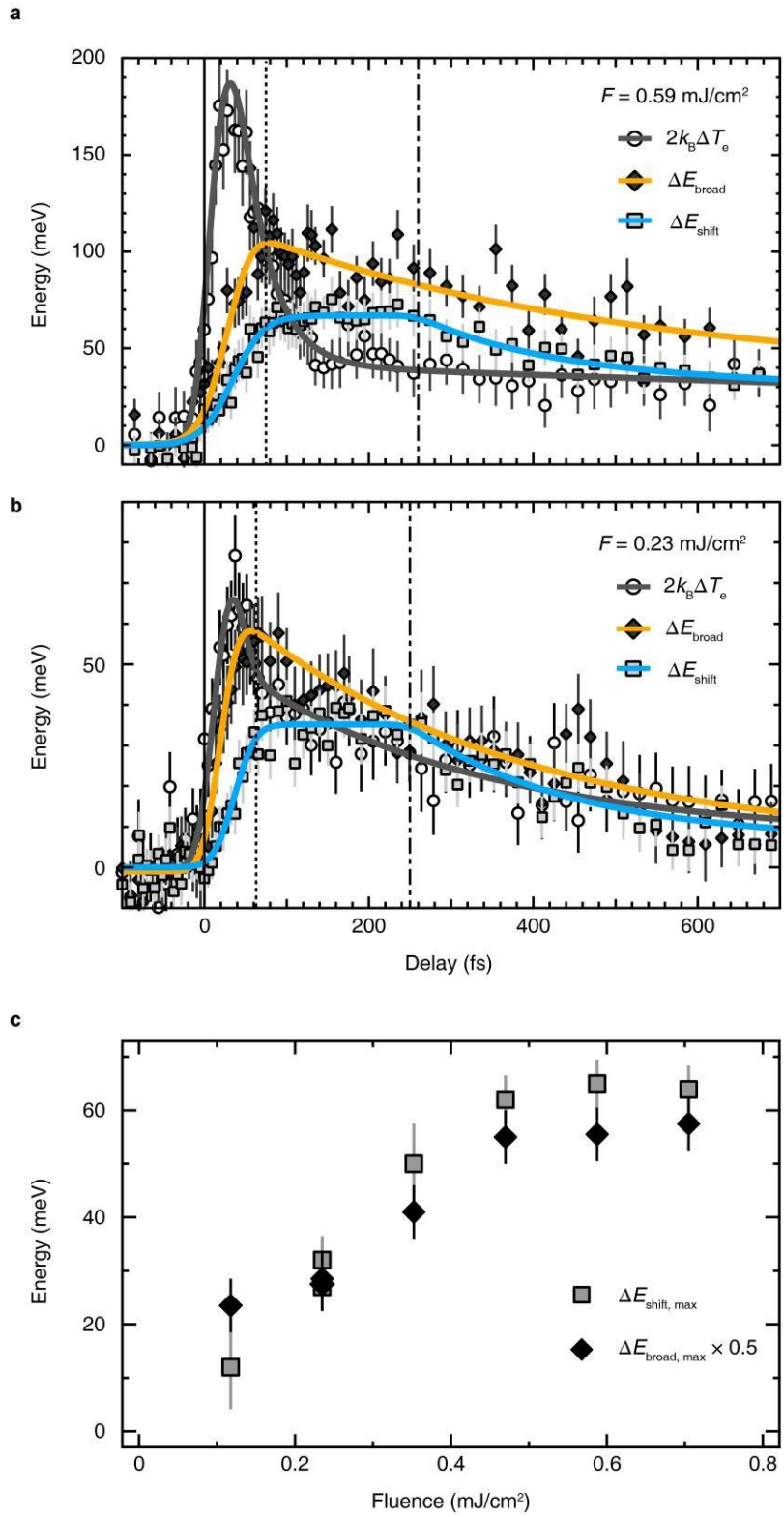